\documentclass[nohyper,11pt,letterpaper]{JHEP3}
\usepackage{graphicx}

\DeclareSymbolFont{AMSa}{U}{msa}{m}{n}
\DeclareSymbolFont{AMSb}{U}{msb}{m}{n}
\let\Box\relax
\DeclareMathSymbol{\Box}{\mathord}{AMSa}{"03}

\def \eqn#1#2{\begin{equation}#2\label{#1}\end{equation}}

\title{Leptogenesis from  Pseudo-Scalar Driven Inflation.}

\author{W. Fischler and  S.Paban\\
Theory Group,Department of Physics \\ University of Texas, Austin,
TX 78712 \\ E-mail: {\tt fischler, paban@zippy.ph.utexas.edu}}

\abstract{We examine recent claims for a considerable amount of leptogenesis, in some inflationary scenarios, through the gravitational anomaly in the lepton number current. We find that when the short distances contributions are properly included the amount of lepton number generated is actually much smaller.  }

\received{???????? ?st, 2000} \accepted{???????? ?th, 1998}
\preprint{UTTG-05-07}

\begin{document}



\section{\bf Introduction}

Explaining the observed baryon asymmetry is a must of any theory beyond the Standard Model. Many candidates have been examined since Sakharov \cite{Sakharov:1967dj} stated the three conditions necessary to generate this asymmetry. In addition, a number of leptogenesis mechanisms   have been explored since it was shown \cite{Fukugita:1986hr}  that the lepton asymmetry can be converted to baryon symmetry  through sphaleron processes, for a recent review  see \cite{Chen:2007fv}. The aim of this short paper is to examine a recent claim of Alexander, Peskin and Sheikh-Jabbari \cite{Alexander:2004us, Alexander:2007qe} for a new leptogenesis mechanism.

The mechanism of Alexander, Peskin and Sheikh-Jabbari( APS-J)   starts  with the gravitational anomaly of the lepton number current \cite{Alvarez-Gaume:1983ig}:

\eqn{gravanomaly}{ \partial_{\mu} \left( \sqrt{- g} \,{J_L}^{\mu} \right) = \frac{3}{ 16 \pi^2 } R \tilde{R} }  
where $ R \tilde{R}= \frac{1}{2} \epsilon^{\alpha \beta \gamma \delta} R_{\alpha \beta \nu \lambda} {R^{ \nu \lambda}}_{\gamma \delta}$  and 
$${J_L}^{\mu}=\sum_i \,  \left(\bar{l}_{L_i} \gamma^{\mu} {l}_{L_i} - \bar{e}_{L_i} \gamma^{\mu} {e}_{L_i} \right)$$ In this expression $l_L$ represents the $SU(2)_L$ lepton doublets  and $e_L$ represents the $SU(2)_L$ charged lepton singlets.  \footnote{Notice  that our expression disagrees with APS-J who write this anomaly equation as $ \partial_{\mu} \left( {J_L}^{\mu} \right) = \frac{3}{ 16 \pi^2 } R \tilde{R}$. } The associated lepton charge

\eqn{leptoncharge}{ Q_L \equiv \int d^3 x \sqrt{-g} {J_L}^0} varies with time as

\eqn{derivativecharge}{ \frac{d Q_L}{dt} = \frac{3}{ 16 \pi^2 }  \, \int  d^3 x  \,\,  R \tilde{R} } 
The key observation of APS-J is that   a deformation of the Hilbert-Einstein action of the form:
\eqn{inflatonRRcoupling} { \int d^4 x  \,\,  \left(\frac{1}{16 \pi G} \sqrt{-g} \,  R + {\cal{F}}(t)\,\, R \tilde{R} \right)}
will  generate a non-vanishing $<R \tilde{R}>$, which in turn will generate a non-vanishing lepton density through the anomaly equation. 
 In their work ${\cal F}(t)$ is a function of the inflaton field $\phi(t)$, assumed to be a pseudo-scalar,  ${\cal F}(t) = F( \phi(t))$. Other effects associated with this particular deformation have been considered in \cite{Lue:1998mq, Jackiw:2003pm, Smith:2007jm}.

The current ${J_L}^{\mu}$ is the low energy lepton current and does not include right handed neutrinos. At energies above where these particles get a mass, it is possible to define a new current that does not suffer from gravitational anomalies. This new current, however, is not interesting for the purpose of generating baryon number, through spahleron transitions,  since the associated charge is explicitly broken by the neutrino mass terms. APS-J's  computation gives
\eqn{RR}{ < R \tilde{R}  > = \frac{ 4 \cal{N} }{\pi^2} \frac{1 }{ a^3} \dot{\phi} H M_P \left( \frac{1}{{M_P}^6} \int \frac{d^3 k}{(2 \pi)^3} k^3  \right) }
where $a(t)$ is the FRW scale factor during inflation,  $H$ is the Hubble constant at that time and $M_P$ is the reduced Planck mass, $M_P= \frac{1}{\sqrt{ 8 \pi G}} $.   This computation assumed

\eqn{F}{{F}(\phi) = \frac{{\cal{N}}}{16 \pi^2 M_P}  \phi} where ${\cal{N}}$  is the number of stringy degrees of freedom propagating in the loops. The string scale has been assumed to be of the same order of $M_P$. The net lepton number density generated through this effect obtained by APS-J is 
\begin{eqnarray}
n & = & \frac{3}{16 \pi^2} \int_{t_i}^{t_e} dt \, \,\,  <R \tilde{R}>  \nonumber \\ \nonumber \\
& = & \frac{ 4 \cal{N} }{3 \pi^2} \sqrt{2 \epsilon} H M_P^2 \left( \frac{1}{{M_P}^6} \int \frac{d^3 k}{(2 \pi)^3} k^3  \right)  \label{n}
\end{eqnarray}  where $n=J^0_L$, $\epsilon= \frac{1}{2} \frac{\dot{\phi}^2}{ ( H M_P)^2} $ is the slow roll inflation parameter, experimentally of order $O(10^{-2})$, $t_i$ and $t_e$  are respectively the times at which inflation started and ended.  Notice that the integral over time is picked at the beginning of the inflationary era, and hence the result is not sensitive to the time at which inflation ends.

This result is very sensitive to the momentum  ultraviolet cut-off,  APS-J chose it to be the mass of the right-handed neutrino, arguing that above this mass the effect will disappear since the current ceases to be anomalous. As we explained above, the leptonic current that is converted through sphalerons to baryon number remains anomalous. Other computations \cite{Lyth:2005jf}  have chosen the scale to be closer to $M_P$, since pseudoscalars are copious in string theory. Although, a higher cut-off will enhance the effect it will also invalidate the use of perturbation theory since more and more higher dimensional operators will contribute significantly to $<R \tilde{R}>$.

How can we then make physical sense out of this result? The divergence found is the result of the ultraviolet behavior of the field modes. These short wavelengths only probe the local geometry around the point where 
$<R \tilde{R}> $ is being computed and should  not sensitive to the global features of the space-time. Yet the expression (\ref{n}) for $n$  is the product of this UV divergent contribution and the horizon scale  $H$ that represents  long distance effects.  We believe that this result reflects an improper renormalization procedure. This puzzling mixing, between long distance effects and the short distance effects,  is reminiscent of finite temperature Quantum Field Theory. 
In the next section, we will explain how a similar issue first appeared in finite temperature QFT and how the puzzle was resolved in that context. Readers familiar with this issue can skip to the next section.

\section{Finite Temperature Quantum Field Theory}

For pedagogical reasons we will briefly review this issue.  The computation of the free energy  for a massive scalar field $\phi$ $(\mbox{mass}=M)$, with a $\lambda \phi^4$ interaction at a large but finite temperature $T>M$, generates at two loops  a term of the form $\lambda T^2\Lambda_{UV}^2$ .

This term involves a mixing between the temperature scale and the UV cut-off scale. The UV dependence of this term is taken care of  by $T=0$ counter-terms. In particular,  at this order in a perturbative expansion, the zero temperature  one loop renormalization of the scalar field's mass squared , insures the absence of this divergent term in the free energy. In what follows we briefly recall this standard procedure with some additional details.

A computation of the two point function in a $\lambda \phi^4$ theory, to first order in $\lambda$,  at finite $T$  \cite{Kapusta:1989tk} , gives:

\eqn{FiniteT}{  \Pi = 12 \lambda T \sum_n \int \frac{ d^3 p}{ ( 2 \pi)^3}  \frac{1}{ \omega^2 +\omega_n^2 } } which displays the announced mixing. In this formula $\omega=(\bar{p}^2 + m^2)^{1/2}$ and $\omega_n= 2 \pi n T$. This expression is yet to be properly renormalized. The sum over all possible frequencies naturally separates into one term that is  temperature independent  $\Pi^{vac}$ ( vacuum piece) and divergent and one term containing the Bose-Einstein distribution, $\Pi^{mat}$ ( matter piece), that is finite and temperature dependent.
\begin{eqnarray*}
\Pi  & = & \Pi^{vac} + \Pi^{mat} \\ \\
\Pi^{vac}  & = & 12 \lambda \int \frac{d^4 p}{ (2 \pi)^4}  \frac{1}{ p_4^2 + {\bar{p}}^2 + m^2} \\ \\
\Pi^{mat} &  = & 12 \lambda \int \frac{d^3 p}{ (2 \pi)^3}  \frac{1}{\omega} \frac{1}{ e^{\beta \omega} -1 } \\
\end{eqnarray*}  

To regulate this divergence the $T=0$  mass counter-term is added to the Lagrangian. The complete renormalized self-energy at $T>0$, at first order in $\lambda$, is

$$\Pi^{ren} = 12 \lambda \int \frac{d^3 p}{ (2 \pi)^3}  \frac{1}{\omega} \frac{1}{ e^{\beta \omega} -1 } $$ 

Inserting this information in a two loop computation of the free energy  leads to an expression for the free energy that  is temperature dependent but void of UV divergences, it doesn't contain the mixing between short distance effects and finite temperature effects.  We do expect an analogous mechanism to be at work in the case of interest.

\section{One-Loop Effective Action}

In order to get a value for $<R\tilde{R}>$, we will compute the one-loop effective action  for the lagrangian given in (\ref{inflatonRRcoupling}) using the background field method. We will treat ${\cal{F}}(t)$ as an external source  and expand the gravitational action around a background $ \hat{g}_{\mu \nu} $, $$ g_{\mu \nu} =  \hat{g}_{\mu \nu} + h_{\mu \nu} $$ Given the symmetries of the inflaton-gravity coupling:

\begin{itemize}
\item diffeomorphism invariance
\item ${\cal{F}}(t)  \rightarrow {\cal{F}}(t) + \mbox{constant}$

\end{itemize}  
together with the fact that ${\cal F}(t) $ is a pseudo-scalar,  the one-loop effective action compatible with these symmetries will be of the form:

\begin{eqnarray} 
 W[{\cal F}] ={M_P}^2  \int \,\, d^4 x  \sqrt{-\hat{g}}   \left\{ \right. & &  \left.  \hat{R} +   a_1 \, ( \partial_{\mu} {\cal F} \, {\hat g} ^{\mu \nu}  \partial_{\nu} {\cal F} ) \,   \right. \nonumber \\
 &  + & \left.  a_2\,  \frac{1}{{M_P}^2} \,{ \hat R}^2 + a_3 \,\frac{1}{{M_P}^2} \, {\hat R}_{\mu \nu} {\hat R}^{\mu \nu}  
 +  ( a_4 \, \frac{{ \hat R }}{{M_P}^2} \,  {\hat g} ^{\mu \nu} +  a_5 \, \frac{ {\hat R} ^{\mu \nu}}{{M_P}^2}) \, ( \partial_{\mu} {\cal F} \, \partial_{\nu} {\cal F} )  \right. \nonumber \\ 
& + & \left.  a_6 \,\frac{1}{{M_P}^2} \, ( \partial_{\mu} {\cal F} \,   {\hat g} ^{\mu \nu}   \partial_{\nu} {\cal F} ) ^2  \, +  a_7 \, \frac{1}{{M_P}^2} \,( \Box {\cal F} ) ^2      +      \ldots       \right\}  \label{Weff}
\end{eqnarray}
where $\Box {\cal F} =\hat{g}^{\mu \nu} D_{\mu} \partial_{\nu} {\cal F}$, and $D_{\mu}$ is the covariant derivative.  It is possible that some of these terms are related by field redefinitions but we won't concern ourselves about this issue. The omitted terms will include higher powers of the fields ${\cal F}$ and  ${\hat{g}}^{\mu \nu}$,  as well as derivatives acting on them.  The coefficients $a_i$ will be  dimensionless. When divergent, the value for the $a_i$ are background independent   \cite{'tHooft:1974bx}. This latter point is rather important, because the renormalization procedure is based  on it. On physical grounds we do expect different backgrounds to have a common ultra violet behavior since the short distance effects are insensitive to global features of the space time. This will not be true of the finite terms which include effects from long wavelengths.   For illustration purposes, in Appendix A we have included the explicit expressions for the graviton propagators in two different backgrounds: flat (\ref{flatpropx})  and de Sitter (\ref{eq:deSitterpropx}). Their dominant short distance behavior agrees by design.  We would expect these two propagators to  start to deviate from each other at distances of the order of  the de Sitter horizon and bigger. This effect is reflected in the expression for the effective action. For example, while in the flat background,  strictly $\hat{g}_{\mu \nu}= \eta_{\mu \nu}$, the action will have only one term proportional to $(\partial{{\cal F}})^2$, in the de Sitter case \footnote{ We are assuming a de Sitter metric of the form $ ds^2 = - dt^2 + e^{2 H t} d{\bar{x}}^2$.}, this term will be multiplied by an infinite series in powers of $ (H/M_P)^2$.

The action  (\ref{inflatonRRcoupling}) is non-renormalizable and the loop computation that we wish to carry out here is only consistent  if it is understood as an effective field theory computation. The procedure is similar to the one followed when one calculates higher loop corrections to the non-linear $\sigma$-model for pions at finite temperature. The only additional challenge is regulating the theory while respecting general covariance.  

Dimensional regularization is well suited for that purpose, however it obscures the mechanism by which the power law divergences cancel out.
A momentum cut-off does violate general covariance, nevertheless the symmetries of the system at hand will dictate the structure of the  counter-terms.

The one-loop corrections to the action will, both in the context of inflation and pions at finite temperature, include terms that involve power law cut-off dependence multiplied by the Hubble constant (in the case for de Sitter) or temperature (in the case for pions).

The challenge as we explained earlier in section 2, is how to disentangle this peculiar mixing of scales.
For example, in the case for de Sitter there will be a contribution to the action among others in the form:

\eqn{term1}{{{{\Lambda_{UV }^4} H^2 }\over{M_P^6}}( \partial_{\mu} {\cal F}  {\hat g} ^{\mu \nu}  \partial_{\nu} {\cal F} ) }where H is the Hubble constant. The resolution  lies in recognizing what the origin for this term is, when considering the effective action in a general gravitational background, and  then evaluating it in the specific geometry at hand, in our case deSitter space. 

There are in general several terms that will contribute. The original terms in the action that leads, when evaluated in de Sitter, to the kinetic terms for ${\cal F}$ proportional to $H^2$, are:

\eqn{term2}{{{{\Lambda_{UV} }^4} \over{M_P^6}}( \partial_{\mu} {\cal F}  {\hat g} ^{\mu \nu}  \partial_{\nu} {\cal F} ) R \;\;\;\;\;\;\;\;\;\;\;\;
{{\Lambda_{UV }^4} \over{M_P^6}}( \partial_{\mu} {\cal F}    \partial_{\nu} {\cal F} ) R^{\mu \nu} }

These terms are taken care by counter-terms and for this purpose the reader can think of the effective action for gravitons in flat space. 

There will then be left over contributions that in the de Sitter case will be proportional to

\eqn{term3}{{{{\Lambda_{UV }^2} H^4}\over{M_P^6}}( \partial_{\mu} {\cal F}  {\hat g} ^{\mu \nu}  \partial_{\nu} {\cal F} ) \;\;\;\
\mbox{and} \;\;\;\;  {{ H^6 }\over{M_P^6}}( \partial_{\mu} {\cal F}  {\hat g} ^{\mu \nu}  \partial_{\nu} {\cal F} ) }

The quadratically divergent term will be taken care of in a manner identical to the preceding discussion, which then leaves over the finite term. This procedure is general.

The astute reader might wonder whether there could be finite contributions  that involves a scale
$ M =X^2/M_{P}$ other than H, with M larger than H. The answer  to that concern is no. Ex absurdo, if this was the case, this would mean that there exists a new mass scale X between the Planck scale and the height  of the potential during inflation. Integrating out this scale will then generate contributions to the action in general and modify substantially the potential for $\cal F$ in particular. By fiat, such contributions to the potential are assumed to be absent.

In addition, the left over skeptic should consider the case of  pions at finite temperature $ T, T<f_{\pi}$.  The same procedure  takes care of  the power law divergences and leaves over an effective action for the pions  that does not contain any terms involving a mass scale other than $T$ and $ f_{\pi}$.

The absence of the power law cut-off dependence in the action as discussed above is automatically realized in dimensional regularization. However the discussion about the absence of a new scale is still necessary to complete the argument 

In dimensional regularization,  the first non-vaninshing contribution of the interaction term (\ref{inflatonRRcoupling}) to the term in the one loop effective action that is quadratic in the field  ${\cal F} $ will contain eight derivatives, since in the UV region  the momentum of the field $\cal F$ is the only scale in the problem. The infinite contribution can be cancelled by adding the following counter term to the lagrangian

\eqn{counterterm}{ \Delta {\cal L} = \left( -  c_1 \frac{1}{\epsilon} + c_{2R} -c_2 \right) \, ( \Box^2({\cal F}))^2}
$c_{2R}$ is the renormalized coupling  and $\epsilon$ is related to the dimension of space time as $d=4-2 \epsilon$.

Likewise, the vertex (\ref{inflatonRRcoupling}) will contribute to terms in the lagrangian of the form  \footnote{The notation is schematic, $R$ stands not only for the scalar curvature but also for the Ricci tensor.} : $ \,\,\,R^n  {\cal F}^2 $, with $n$ at one loop ranging from $1$ to $3$.   Following a similar power counting procedure \footnote{ There are four derivative on each vertex}  the first finite contribution will be schematically of the form $ R^n  \,  \partial^{8-2 n} ({\cal F}  \, {\cal F}) $.  In agreement with the discussion above, for example,  this interaction will not produce terms of the form (\ref{term2}) but rather

\eqn{term2d}{\frac{1}{M_P^6} \, ( \partial_{\mu} \Box {\cal F}  {\hat g} ^{\mu \nu}  \partial_{\nu} \Box  {\cal F} ) R \;\;\;\;\;\;\;\;\;\;\;\;
{1 \over{M_P^6}}( \partial_{\mu} \Box {\cal F}    \partial_{\nu} \Box {\cal F} ) R^{\mu \nu} }

The ultimate goal  is to determine which of these terms makes the biggest contribution to:

\eqn{finalRRtilde}{ < R \tilde{R}> = \frac{ \delta W[{\cal F}]}{ \delta {\cal F} (t)} }
 
The part of the renormalized effective action that depends on the field ${\cal F}(t)$, in a Friedman-Roberston-Walker background, can be written as:

\begin{eqnarray}
 W[{\cal F}] = \int \,\, d^4 x \,\, a^3(t) \frac{1}{{M_P}^4} \,  \left\{  \right. & &   \, \alpha_1 \left. \left( \frac{d^4{\cal F}(t)}{dt^4} \right)^2 + \alpha_2 \, R  \left( \frac{d^3{\cal F}(t)}{dt^3} \right)^2  \right. \nonumber \\
& + & \left. \alpha_3 \, R^2  \left( \frac{d^2{\cal F}(t)}{dt^2} \right)^2 + \alpha_4 \,  R^3  \left( \frac{d{\cal F}(t)}{dt} \right)^2   +  \ldots \right\}  \label{WFRW} 
\end{eqnarray}

\section{Lepton number density generated by the inflaton background}
In this and the remaining computations we should restrict the background to be a space-time metric of constant curvature: Minkowski or de Sitter. From (\ref{finalRRtilde}),
 $$ <R \tilde{R}> = \frac{1}{{M_P}^4} \, \sum_{n=0}^{3}  \frac{ d^{4-n} }{ dt^{4-n} } \, \left[  (-1)^n a^3(t) \,   \alpha_{n+1}  \,  R^{n} \, \frac{d^{4-n}{\cal  F}}{d t^{4-n}} \right]   \label{finaldeSitterRR} $$

To compare this result with the one obtained by APS-J we need to determine which of the terms in the expression above will give the biggest contribution. Assuming ${\cal F}(t)$ to be of the form (\ref{F}), and assuming a slow roll approximation, that is $\frac{d^n \phi}{dt^n} \ll ( \frac{d \phi}{dt} )^n$, the dominant contribution will be 

\eqn{veryfinalRR}{ <R\tilde{R}> = \frac{1}{{M_P}^4}  \frac{d}{dt}  \left[ \, a^3(t)  \,  \alpha_{4}  \,  R^{3} \, \frac{d{\cal  F}}{d t}   \right] }The different dependence in the scale factor between this expression and  (\ref{RR}) is due to the different definitions of the current.
 The contribution to the lepton number density, $n=J^0_L$,  (\ref{n}) is:

\eqn{finaln}{\Delta( a^3(t) \, n) = \frac{3}{16 \pi^2} \, \int_{t_i}^{t_e} dt  \, <R \tilde{R}>  }

\eqn{n1}{ n(t_e)=  \, \frac{{\cal N}}{(16 \pi^2)} \, \frac{5184 }{ (16 \pi^2) } \, \alpha_4 \, \sqrt{2 \epsilon} \, \frac{H^7}{ {M_P}^4}  \left( 1 - e^{-3N_e} \right) } 
where $N_e$ represents the number of e-foldings. This result should be compared with the entropy density at the end of reheating. Assuming that reheat is instantaneous \cite{Alexander:2004us}, that is the reheating temperature is given by the equality  $\rho= 3 H^2 M_P^2= \pi^2 g_* T^4 /30$ , the entropy is
$s= 2.3 \, g_*^{1/4} ( H M_P)^{3/2}$. The predicted $n/s$ ratio will  then be:

\eqn{nsratio}{ \frac{n}{s} = 2254 \,  \left( \frac{1}{  3 ^{11} g_* } \right)^{1/4}  \, \alpha_4   \, \sqrt{2 \epsilon}  \frac{{\cal N}}{ ( 16 \pi^2)^2} \left( \frac{ \Lambda^{1/4} }{ M_P} \right)^{11}}
where $H=  \sqrt{\Lambda} / {( { M_P} \sqrt{3})}$. The observed baryon density  ratio is  $n/s= 2.4 \times 10^{-10}$. WMAP \cite{Spergel:2006hy} has put an upper bound on the scale of inflation to be $\Lambda^{1/4} < 2 \times 10^{16} \mbox{GeV}$. With these asumptions:

\eqn{nnum}{ \frac{n}{s} = 1.56  \times 10^{-25} \, \left( \frac{\alpha_4}{1} \right) \, \left( \frac{ \sqrt{2 \epsilon}}{ 10^{-1}} \right)  \left( \frac{{\cal N}}{100} \right)  \, \left( \frac{100}{g_*} \right)^{1/4} \, \left( \frac{ \Lambda^{1/4}}{{ 2 \times 10^{16} \mbox{GeV}}} \right)^{11}}

\section{\bf Conclusions}

A deformation of the Einstein-Hilbert action of the form (\ref{inflatonRRcoupling}) will change, during a pseudo-scalar driven inflation, the total lepton number but in an amount that is too small to explain the observed net baryon number, even when the free parameters are stretched to the most favorable limits compatible with experimental bounds.

\section{Acknowledgments}

Paban thanks Arkady Vainshtein and Richard Woodard for helpful discussions and the Aspen Center for Physics for its hospitality during the completion of this work. This material is based upon work supported by the National Science Foundation under Grant No. PHY-0455649.

\section{Appendix  A} 
 
In our notation perturbation theory is derived from the expansion

\eqn{metricpert}{ g_{\mu \nu} =  \hat{g}_{\mu \nu} + h_{\mu \nu}  }
 where $  \hat{g}_{\mu \nu} $ is the background metric.  In this weak field approximation
\eqn{Riemann}{ R_{\lambda \mu \nu \kappa}= \frac{1}{2} \left\{ \frac{ \partial h_{\lambda \nu}}{ \partial x^{\mu} \partial x^{\kappa}}  -  \frac{ \partial h_{\mu \nu}}{ \partial x^{\lambda } \partial x^{\kappa}}  -  \frac{ \partial h_{\lambda  \kappa}}{ \partial x^{\nu} \partial x^{\mu}}  +  \frac{ \partial h_{\mu \kappa}}{ \partial x^{\nu} \partial x^{\lambda}}  \right\} }

In the Harmonic gauge, the form of the graviton propagator around the Minkowski background , $  \hat{g}_{\mu \nu} =\eta_{\mu \nu}=(-,+,+,+)  $, is
 
\eqn{flatprop}{  P_{\mu \nu, \alpha \beta} (k)= \frac{1}{2 {M_P}^2} \frac{1}{k^2- i \epsilon} \left({ \eta_{\mu \alpha} \eta_{\nu \beta}+   \eta_{\mu \beta} \eta_{\nu \alpha}-  \eta_{\mu \nu} \eta_{\alpha \beta}}\right)} where $M_P= \frac{1}{\sqrt{ 8 \pi G}} $ is the reduced Planck mass. 

In coordinate space it can be written as:

\eqn{flatpropx}{D_{\mu \nu, \alpha \beta} (x,x')= \frac{1}{ {M_P}^2} \frac{i}{ 4 \pi^2}  \frac{1}{\sigma(x, x')^2- i \epsilon} \left({ \eta_{\mu \alpha} \eta_{\nu \beta}+   \eta_{\mu \beta} \eta_{\nu \alpha}-  \eta_{\mu \nu} \eta_{\alpha \beta}}\right)}  where $ \sigma(x,x')^2 \equiv \eta_{\mu \nu} ( x-x')^{\mu} ( x-x')^{\nu} $.

We will assume the Friedman-Robertson-Walker metric for de Sitter space:

$$ ds^2 = - (d t)^2 + a(t)^2 \delta_{ij} dx^i dx^j $$  where $a(t)= e^{ H t}$. As always $ H= \sqrt{\Lambda} / {( { M_P} \sqrt{3})}$. The expressions that will follow are more easily derived in conformal time. 

$$ \eta= \frac{1}{H a } = \frac{1}{H} e^{- Ht} $$ Also in the Harmonic gauge, the graviton propagator following \cite{Tsamis:1992zt} is given by,

\begin{eqnarray}
D^{dS}_{\mu \nu, \alpha \beta} (x,x') & = &  \frac{1}{ {M_P}^2} \frac{i}{4 \pi^2} \frac{a(\eta) a(\eta')}{\lambda(x,x')^2- i \epsilon} \left({ \eta_{\mu \alpha} \eta_{\nu \beta}+   \eta_{\mu \beta} \eta_{\nu \alpha}-  \eta_{\mu \nu} \eta_{\alpha \beta}}\right)  \nonumber \\ \nonumber \\
&  - &  \frac{1}{{ M_P}^2} \frac{1}{ 4 \pi^2} \frac{ a(\eta) a(\eta') }{ \eta \eta'} \ln [ H^2 \lambda(x,x')^2 + i \epsilon] \left(  \bar{\delta}_{\mu \alpha} \bar{\delta}_{ \nu \beta} +  \bar{\delta}_{\mu \beta} \bar{\delta}_{\nu \alpha} -  \eta_{\mu \nu} \eta_{\alpha \beta} \right) \nonumber \\
\label{eq:deSitterpropx}
\end{eqnarray} where $\lambda(x,x')^2 \equiv -(\eta-\eta')^2 +(\bar{x}-\bar{x'})^2$ and a bar over  the tensor $\delta_{\alpha \beta} $ indicates the suppression of the zero components. 
This propagator is derived with the constraint that it match the flat space propagator in the limit of $\mu/(1/H) \rightarrow 0$. More precisely, the part of the  propagator  that is proportional to $\left({ \eta_{\mu \alpha} \eta_{\nu \beta}+   \eta_{\mu \beta} \eta_{\nu \alpha}-  \eta_{\mu \nu} \eta_{\alpha \beta}}\right)$ must be proportional to $(1-z)^{-1}$ as $\mu H \rightarrow 0$. The other terms must either be finite or diverge more slowly than $(1-z)^{-1}$. The variable $z$ is related to the geodesic distance between two points through the relation:

\eqn{z}{z(x,x')= \cos \left( \frac{ \mu H}{2} \right) ^2= 1 +  \frac{1}{4 \eta \eta' } \lambda(x,x')}
 The de Sitter propagator  (\ref {eq:deSitterpropx}) reduces to the flat propagator (\ref{flatprop}) in the limit $H \rightarrow 0$.

%

\end{document}